\documentstyle [11pt]{article}
\def\journal#1#2#3#4{\  {#1} \ {\bf #2}, {#3}\  ({#4})}

\def\AnnPhys{\journal{Ann.\ Phys.}}

\def\ibid{\journal{\em ibid.}}

\def\NPB{\journal{Nucl.\ Phys.\ {\bf B}}}

\def\PhysRev{\journal{Phys.\ Rev.}}

\def\PRD{\journal{Phys.\ Rev.\ {\bf D}}}

\def\SovJNuclPhys{\journal{Sov.\ J.\ Nucl.\ Phys.}}

\begin{document}

\newcommand{\gn}{\mbox{$\gamma_{\stackrel{}{5}}$}}
\newcommand{\adag}{a^{\dagger}_{p,s}}
\newcommand{\atildedag}{\tilde{a}^{\dagger}_{-p,s}}
\newcommand{\bdag}{b^{\dagger}_{-p,s}}
\newcommand{\btildedag}{\tilde{b}^{\dagger}_{-p,s}}
\newcommand{\apsbeta}{a^{\beta}_{p,s}}
\newcommand{\apsbetadag}{a_{-p,s}^{\beta\dagger}}
\newcommand{\adagbdag}{a^{\dagger}_{p,s} b^{\dagger}_{-p,s}}
\newcommand{\aps}{a^{}_{p,s}}
\newcommand{\bps}{b^{}_{-p,s}}
\newcommand{\bpsbeta}{b^{\beta}_{p,s}}
\newcommand{\bpsbetadag}{b_{-p,s}^{\beta\dagger}}
\newcommand{\Adag}{A^{\dagger}_{p,s}}
\newcommand{\Bdag}{B^{\dagger}_{-p,s}}
\newcommand{\Aps}{A^{}_{p,s}}
\newcommand{\Apsbeta}{A^{\beta}_{p,s}}
\newcommand{\Apsbetadag}{A^{\beta\dagger}_{-p,s}}
\newcommand{\Bps}{B^{}_{p,s}}
\newcommand{\Bpsbeta}{B^{\beta}_{p,s}}
\newcommand{\Bpsbetadag}{B^{\beta\dagger}_{-p,s}}
\newcommand{\ApL}{A^{}_{p,L}}
\newcommand{\BpL}{B^{}_{-p,L}}
\newcommand{\ApR}{A^{}_{p,R}}
\newcommand{\BpR}{B^{}_{-p,R}}
\newcommand{\apL}{a^{}_{p,L}}
\newcommand{\bpL}{b^{}_{-p,L}}
\newcommand{\apR}{a^{}_{p,R}}
\newcommand{\bpR}{b^{}_{-p,R}}
\newcommand{\AdagL}{A^{\dagger}_{p,L}}
\newcommand{\AdagR}{A^{\dagger}_{p,R}}
\newcommand{\BdagL}{B^{\dagger}_{-p,L}}
\newcommand{\BdagR}{B^{\dagger}_{-p,R}}
\newcommand{\adagL}{a^{\dagger}_{p,L}}
\newcommand{\adagR}{a^{\dagger}_{p,R}}
\newcommand{\bdagL}{b^{\dagger}_{-p,L}}
\newcommand{\bdagR}{b^{\dagger}_{-p,R}}
\newcommand{\eps}{\epsilon}
\newcommand{\gnplus}{\gamma \cdot n_{_{+}}}
\newcommand{\gnminus}{\gamma \cdot n_{_{-}}}
\newcommand{\gnplusdef}{\left( \vec{\gamma} \cdot \hat{n}-\gamma_o \right)}
\newcommand{\gnminusdef}{\left( \vec{\gamma} \cdot \hat{n}+\gamma_o \right)}
\newcommand{\abab}{a^{\dagger}_{p,L}\,b^{\dagger}_{-p,L}\,a^{\dagger}_{p,R}
                   \,b^{\dagger}_{-p,R}}
\newcommand{\alphai}{\alpha_{i}}
\newcommand{\limit}{\lim_{\Lambda^2 \rightarrow \infty}}
\newcommand{\p}{\vec{p}, p_o}
\newcommand{\poprime}{p_o^{\prime}}
\newcommand{\prodps}{\prod_{p,s}}
\newcommand{\prodp}{\prod_{p}}
\newcommand{\psibar}{\bar{\psi}}
\newcommand{\psibarpsi}{ < \bar{\psi} \, \psi
            > }
\newcommand{\PsibarPsi}{ < \bar{\Psi} \, \Psi
            > }
\newcommand{\psibeta}{\psi^{}_{\beta}}
\newcommand{\psibarbeta}{\bar{\psi}_{\beta}}
\newcommand{\psidag}{\psi^{\dagger}}
\newcommand{\psidagbeta}{\psi^{\dagger}_{\beta}}
\newcommand{\psiL}{\psi_{_{L}}}
\newcommand{\psiR}{\psi_{_{R}}}
\newcommand{\Q}{Q_{_{5}}}
\newcommand{\Qa}{Q_{_{5}}^{a}}
\newcommand{\Qbeta}{Q_{5}^{\beta}}
\newcommand{\qqbar}{q\bar{q}}
\newcommand{\sumps}{\sum_{p,s}}
\newcommand{\thetap}{\theta_{p}}
\newcommand{\costhetap}{\cos{\thetap}}
\newcommand{\sinthetap}{\sin{\thetap}}
\newcommand{\thetaset}{\{ \thetap  \}}
\newcommand{\thetapi}{\thetap{}_{i}}
\newcommand{\Tomega}{\frac{\Tprime}{\omega}}
\newcommand{\pomega}{\frac{p}{\omega}}
\newcommand{\Tprime}{T'}
\newcommand{\Tprimesq}{T^{'2}}
\newcommand{\vac}{| vac \rangle}
\newcommand{\vacbeta}{| vac \rangle_{_{\beta}}}
\newcommand{\x}{\vec{x},t}
\newcommand{\xPrime}{\vec{x} - \hat{n} (t - t' ), t'}
\newcommand{\xPrimet}{\vec{x} + \hat{n} (t - t'), t'}
\newcommand{\y}{\vec{y}, y_o}

\vspace*{-.15in}

\hspace*{\fill}\fbox{CCNY-HEP-94-8}

\begin{center}
{\Large {\bf Spacetime Quantization of BPFTW Action: \\
Spacelike Plasmon Cut \& New Phase of the
	Thermal Vacuum\fnsymbol{footnote}\footnote
{\samepage \sl
\noindent \parbox[t]{138mm}{ \noindent
                       This work has been supported in part by a
                       grant from PSC-BHE of CUNY.
                           }

}
}
}\\
\baselineskip 5mm
\ \\
Ngee-Pong Chang (npccc@cunyvm.cuny.edu)\\
Department of Physics\\
City College \& The Graduate School of City University of New York\\
New York, N.Y. 10031\\
\  \\
Sept 24, 1994  \\
\end{center}

\vspace*{-.15in}

\noindent\hspace*{\fill}\parbox[t]{5.5in}{
        \hspace*{\fill}{\bf Abstract}\hspace*{\fill} \\
        {\em
	The properties of quark propagation through a hot medium
	are summarized by the BPFTW effective action.
	The fermion thermal propagator shows a pseudo-Lorentz
	invariant particle pole as well as a spacelike
	cut.
	In an earlier paper, we have performed the explicit
	quantization of the action in momentum space, and
	showed how a canonical Dirac field of mass $\Tprime$
	arises.  \\

	In this paper, we perform the quantization in coordinate
	space.  In the process, we relate the
	spacelike plasmon cut in the propagator to the homogeneous
	solutions of the local equation of motion for auxiliary fields. \\

	Our quantization shows how the spacelike cut produces a $90^{o}$
	phase factor in the thermal vacuum at high $T$.
	This phase factor is responsible for the vanishing
	of $ \psibarpsi $ at high $T$. \\

        }
                              }\hspace*{\fill} \\

\begin{flushleft}
PACS: 12.38 Aw, 11.10.wx, 11.15.Ex, 11.30.Rd
\end{flushleft}


\section{Introduction}

	In an earlier paper$\cite{Chang-xc}$, we have studied the
	canonical quantization of the BPFTW$\cite{BP}$ action for
	the propagation of a massless fermion through the hot
	environ$\cite{Weldon-Klimov}$.
	The propagator
	$ <T\left( \psi (\vec{x}, t) \psibar (\vec{y}, t')
			\right) >_{_{\beta}} $ in spite of its
	apparent chiral symmetry shows a pseudo-Lorentz invariant
	particle pole$\cite{Donoghue-Chang-hiT-Barton}$ at
	$p_o = \omega \equiv \sqrt{ \vec{p} \,{}^2 + \Tprimesq }$.
	This pole is described by the canonical Dirac field,
\begin{equation}
	\Psi (\x)  = \frac{1}{\sqrt{V}}  \sum_{p,s}
		{\rm e}^{i \vec{p} \cdot \vec{x}}
		\left\{ U_{p,s} A_{p,s}^{} {\rm e}^{-i \omega t}
			+ V_{-p,s} B_{-p,s}^{\dagger} {\rm e}^{+ i \omega t}
			\right\}		\label{eq-Psi-canon-exp}
\end{equation}
	where $\Aps, \Bps$ are the annihilation operators for a
	particle of mass $\Tprime$, and $U_{ps}, V_{-p,s}$ are the
	massive Dirac spinors$\cite{Chang-hisig}$
 	In addition, the thermal fermion propagator shows
	a pair of parallel space like cuts, just above and below the real
	axis and running from $p_o = - |\vec{p}|$ to $p_o = |\vec{p}|$.

	This pair of spacelike cuts is attributed to the hot plasma state.
	But how exactly is it related to the thermal vacuum in the
	plasma state?
	In this paper we investigate the origin of the spacelike
	cut in terms of the space-time quantization of the BPFTW action.

\section{BPFTW Action}

	The BPFTW action$\cite{BP}$ for the propagation of the fermion through
	a hot environ takes the form
\begin{equation}
   {\cal L}_{\rm eff} = - \psibar \gamma_{\mu} \partial^{\mu}
                          \psi
                       - \frac{T^{'2}}{2\;\;} \, \psibar
                         \left<
                       \frac{\gamma_o - \vec{\gamma} \cdot \hat{n} }
                       {D_o + \hat{n} \cdot \vec{D} }
                         \right> \psi          		\label{eq-BP-action}
\end{equation}
	where
\begin{equation}
	T^{'2}  = \frac{\textstyle g_r^2 C_f}{\textstyle 4} T^2
\end{equation}
	and $C_f$ is the Casimir invariant equal to $( N^2 - 1)/( 2 N )$
	for $SU(N)$ group.
        The angular bracket denotes an average over the orientation
        $\hat{n}$.

	Eq.(\ref{eq-BP-action}) is a nonlocal action in spacetime.  If we
	suppress the gluon fields and concentrate on the fermion
	sector of the effective action, then the
	nonlocality is a lightlike separation between the two fermion fields
\begin{eqnarray}
        {\cal L}_{\rm eff} &=&
                  - \psibar \gamma_{\mu} \partial^{\mu} \psi
                  + \frac{\Tprimesq}{8\;\; } \int dt' \;
			\eps(t-t') \;\left\{ \right.
						\nonumber \\
             & &  \;\;\;\; < \;\psibar (x)
		  \gnplusdef \psi( x_{_{P}} ) \;>
             		\;-\; < \;\psibar (x)
                  \gnminusdef \psi( x_{_{P'}} )  \;>
                              \left. \;\right\}     \label{BP}
\end{eqnarray}
	where $x_{_{P}}$ and $x_{_{P'}}$ refer to the pair of conjugate
	points lightlike separated from $(\x)$
\begin{equation}
	x_{_{P}}  = (\xPrime), \;\;\;\; x_{_{P'}} = ( \xPrimet)
\end{equation}
	The nonlocal Euler-Lagrange equation of motion for the action is
\begin{eqnarray}
	\gamma_{\mu}  \partial^{\mu} \;\psi (\x) &=&
		\;\;\;\frac{\Tprimesq}{8\;\;}
                  \int dt^{\prime} \;\eps (t - t') \;\;< \;\gnplusdef
                  \psi(\xPrime) \;>    	\nonumber \\
          & & 	- \frac{\Tprimesq}{8\;\;}
		  \int dt' \;\eps (t - t') \;\;< \;\gnminusdef
                  \psi(\xPrimet) \;> 		\label{eq-nonlocal}
\end{eqnarray}

	Following Weldon$\cite{Weldon-BP}$, we may replace the nonlocal
	action by a local one involving auxiliary fields$\cite{notation}$
\begin{eqnarray}
	\gamma_{\mu} \partial^{\mu} \; \psi (\x) &=&  \frac{\Tprime}{2\;\;}
			< \;\gnplus \;\;\chi_{_{_{+}}} \;>
			\;\;-\;\; \frac{\Tprime}{2\;\;}
			< \;\gnminus \;\chi_{_{_{-}}}
			\;>  			\label{eq-local-1} \\
	n_{_{ +, \mu }} \partial^{\mu} \;\chi_{_{_{+}}} &=&
			\frac{\Tprime}{2\;\;}  \psi (\x)
						\label{eq-local-2} \\
	n_{_{ -, \mu }} \partial^{\mu} \;\chi_{_{_{-}}} &=&
			\frac{\Tprime}{2\;\;}
				\psi (\x) 	\label{eq-local-3}
\end{eqnarray}
	where we have introduced the two conjugate lightlike 4-vectors
\begin{eqnarray}
	n_{_{+, \mu}} &\equiv& ( \hat{n}, \;\;1 )  \\
	n_{_{-, \mu}} &\equiv& ( \hat{n}, -   1 )
\end{eqnarray}
	so that
\begin{eqnarray}
	\gnplus		&=& \gnplusdef  \\
	\gnminus	&=& \gnminusdef
\end{eqnarray}
	and the new local effective Lagrangian takes the form
\begin{eqnarray}
	{\cal L}_{\rm eff} =  - \psibar \gamma_{\mu} \partial^{\mu} \psi
		& & 	+ \;\;\;\;\frac{\Tprime}{2\;\;} \psibar \left<
			\gnplus \;\chi_{_{+}}  \right>
			\;\;\;\;\;-\;\;\; \frac{\Tprime}{2\;\;} \psibar
			\left< \gnminus
			\;\chi_{_{-}}  \right> \nonumber \\
		& & 	- \;\;\;\;\frac{\Tprime}{2\;\;} \left<
			\bar{\chi}_{_{+}} \gnplus \;\psi \right>
			\;\;\;\;\;+\;\;\; \frac{\Tprime}{2\;\;}
			\left< \bar{\chi}_{_{-}}
			\gnminus \;\psi \right>  \nonumber \\
		& & 	+  \left< \bar{\chi}_{_{+}} \;\gnplus \;
			n_{_{+}} \cdot \partial \;\chi_{_{+}} \right>
			\;-\; \left< \bar{\chi}_{_{-}} \; \gnminus \;
			n_{_{-}} \cdot \partial \;\chi_{_{-}}
			\right>			\label{eq-L-eff-local}
\end{eqnarray}

	Note that apart from the equations of motion, the auxiliary
	fields, $\chi_{_{\pm}}$, do not satisfy any other constraints.
	This is unlike in the case of Weldon$\cite{Weldon-BP}$, where his
	auxiliary field, $\phi^{Weldon}$, satisfies the constraint
\begin{equation}
	\gnplus \gamma_{o}  \;\;\phi^{Weldon} = 2 \;\;\phi^{Weldon}
\end{equation}
	beyond the equation of motion
\begin{equation}
	i n_{_{+ \mu}} \partial^{\mu} \;\phi^{Weldon} (x)  =
			\frac{\Tprime}{2\;\;} \;
			\gnplus \;\psi (x)
\end{equation}
	In addition, Weldon does not distinguish between the two
	light-cones $n_{_{+ \mu}}$ and $n_{_{- \mu}}$.
	The relation between our auxiliary field $\chi_{_{+}}$ and
	that of Weldon is
\begin{equation}
	\phi^{Weldon} (x)  =  -i \;\gnplus \;\chi_{_{+}} (x)
\end{equation}
	Written in this form, the reason for the constraint on Weldon's
	field becomes clear.  It comes about because of the nilpotent
	operator $\left(  \vec{\gamma} \cdot \hat{n} - \gamma_{o} \right)$
	acting on $\chi_{_{+}}$,
\begin{equation}
	\left( \gnplus \right)^{2} = 0
\end{equation}
	The advantage of the coupled set of local equations
	of motion (\ref{eq-local-1}-\ref{eq-local-3}) is that it gives us
	an insight into the role of the homogeneous solutions to
	eq.(\ref{eq-local-2}) and eq.(\ref{eq-local-3}), which
	otherwise would not be transparent with the nonlocal equation
	of motion (\ref{eq-nonlocal}).
	These homogeneous solutions turn out to be related to the spacelike
	cuts of the thermal fermion propagator.

\section{Analytic Properties of Thermal Fermion Propagator}
	The BPFTW action leads to the thermal fermion
	propagator$\cite{Weldon-hole}$
\begin{equation}
	< T( \psi (x) \bar{\psi} (y) )>_{_{\beta}}
	=	\frac{1}{i} \;\int \frac{d^4 p}{(2\pi)^4}
		\;\;\frac{- i \vec{\gamma} \cdot \vec{p}
		( 1 - \frac{\Tprimesq}{2\;\;} a )
		      + i \gamma_o p_o ( 1 - \frac{\Tprimesq}{2\;\;} b ) }
		{p^2 - p_o^2 + \Tprimesq - i \eps + \frac{T' {}^4}{4\;\;}
		\left(  p^2 a^2 - p_o^2 b^2 \right)}
\end{equation}
	where $a, b$ are the functions
\begin{eqnarray}
	a &=&   \frac{p_o}{2 p^3} \ln{\left| \frac{ p_o + p }
			{ p_o - p } \right| }
		- \frac{1}{p^2}  \\
	b &=&   \frac{1}{2p_o p} \ln{\left| \frac{ p_o + p }
			{ p_o - p } \right| }
\end{eqnarray}
	chosen to be real along the entire real $p_o$ axis, and
	$p$ denotes the magnitude of $\vec{p}$.

	For $t > 0$, we have
\begin{eqnarray}
	< T( \psi(x) \bar{\psi}(0) ) >_{_{\beta}} &=& < \psi(x)
		\bar{\psi} (0) >  \\
	&=&	\;\;\; \int \frac{d^3 p}{ (2\pi)^3 } \;
		{\rm e}^{i \vec{p} \cdot \vec{x}} \;\left\{
		Z_{p} \frac{-i \vec{\gamma}
			\cdot \vec{p} + i \gamma_o \omega }{2 \omega} \;
		{\rm e}^{ - i \omega t} \right.\\
	& &	- \left. \frac{\Tprimesq}{8\;\;} \;
		\int_{-p}^{p} \,\frac{dp_o'}{p^3}  \;
		\frac{i \vec{\gamma} \cdot \vec{p} p_o'
		- i \gamma_o p^2}{p^2 - p_o^2 + \Tprimesq}   \;
		{\rm e}^{- i p_o' t} \right\}
		+ O (T'^4)			\label{eq-spacelike-cut}
\end{eqnarray}
	Here, we have performed the contour integration in $p_o$-plane
	and isolated the pole as well as the cut contributions for
	$t > 0$.  In doing so, we have dropped an $O(T^{'4})$ term
	along the spacelike cut.

	Note that the spinor structure of the massive particle pole term
	has the (unusual) feature of being manifestly chiral invariant.
	The wave function renormalization constant $Z_p$ at the pole
	is given by
\begin{equation}
	Z_{p} = 1 - \frac{\Tprimesq}{4p^2} \left( \ln{ \frac{4p^2}{\Tprimesq}}
					- 1 \right)
\end{equation}

\section{Relation between $\psi$ and $\Psi$}
	Based on the earlier work$\cite{Chang-xc}$, the field $\psi$ is
	related to the canonical $\Psi$ by the expansion
\begin{equation}
	\psi (x) = \Psi (x) + \frac{\Tprime}{8\;\;} \int dt' \eps (t-t')
			\left( \;< \;\gnplus \Psi ( x_{_{P}} ) \;>
			\;\;-\;\; < \;\gnminus \Psi ( x_{_{P'}}) \;> \right)
			+ \ldots		\label{eq-psi-expand-1}
\end{equation}
	This expansion correctly describes the thermal fermion propagator
	accurately to order $\Tprime$.

	To go beyond the $O(\Tprime)$ term in the expansion, we go
	back to eq.(\ref{eq-local-2},\ref{eq-local-3})
	and note the homogeneous solutions
\begin{eqnarray}
	\chi_{_{+}}^{s} (\x) &=& \frac{1}{\sqrt{V}} \sum_{p,s}
		\left\{   F_{_{+}} ( \hat{n}, p) \; U_{p,s}  \;\Aps
			+ G_{_{+}} ( \hat{n}, p) \; V_{-p,s} \;\Bdag
		\right\} {\rm e}^{i \vec{p} \cdot \vec{x} - i \hat{n} \cdot
		\vec{p} t}  \\
	\chi_{_{-}}^{s} (\x) &=& \frac{1}{\sqrt{V}} \sum_{p,s}
		\left\{   F_{_{-}} ( \hat{n}, p) \; U_{p,s}  \;\Aps
			+ G_{_{-}} ( \hat{n}, p) \; V_{-p,s} \;\Bdag
		\right\} {\rm e}^{i \vec{p} \cdot \vec{x} + i \hat{n} \cdot
		\vec{p} t}
\end{eqnarray}
	where $\Aps$ and $\Bps$ are the particle annihilation operators of
	the massive canonical Dirac field of eq.(\ref{eq-Psi-canon-exp})
	and the $F_{_{\pm}}$ and $G_{_{\pm}}$ are Dirac matrix functions
	of $\hat{n}, \vec{p}, p_o$.  They are to be determined so that
	they reproduce the location, magnitude and phase of the spacelike
	cuts in the thermal fermion propagator.

	These homogeneous solutions in turn contribute to the expansion
	for $\psi$ through eq.(\ref{eq-local-1}), so that
	eq.(\ref{eq-psi-expand-1}) now becomes
\begin{equation}
	\psi (x) = \Psi (x) + \frac{\Tprime}{8\;\;} \int dt'
			\eps(t-t') \left( < \gnplus
			\Psi (x_{_{P}})> - < \gnminus \Psi (x_{_{P'}})>
			\right)
			+ \Psi^{s} (\x)		\label{eq-psi-Psi-exp}
\end{equation}
	with
\begin{equation}
	\Psi^{s} (x)	=   i  \frac{\Tprime}{2\;\;} \;\int d^4 y \left\{
			\;< S^{-1} (x-y) \gnplus \;\chi_{_{+}}^{s} (y) >
			- < S^{-1} (x-y) \gnminus \;\chi_{_{-}}^{s} (y) >
			\right\}
\end{equation}
	where $S^{-1}$ is the inverse propagator satisfying the property
\begin{equation}
	\left( \gamma \cdot \frac{\partial}{\partial x} + \Tprime\right)
		\; S^{-1} (x-y)
		\;=\; - i \; \delta^{4} (x-y)
\end{equation}

	The Dirac matrix functions, $F_{_{\pm}}$ and $G_{_{\pm}}$, have
	the representation
\begin{eqnarray}
	F_{_{\pm}} ( \hat{n}, \vec{p}, p_o ) &=&
		\pm \;\left( \vec{\gamma} \cdot \hat{n}
		\;\pm\; \gamma_o \,\right)
		\,f_{_{1}} \;\; \pm \;\; \left( \vec{\gamma} \cdot \vec{p}
		\;\mp\; \hat{n} \cdot \vec{p} \;\gamma_o \,\right)
		\,f_{_{2}} 			\label{eq-F-rep} \\
	G_{_{\pm}} ( \hat{n}, \vec{p}, p_o ) &=&
		\pm \;\left( \vec{\gamma} \cdot \hat{n}
		\;\pm\; \gamma_o \,\right)
		\,g_{_{1}} \;\; \pm \;\; \left( \vec{\gamma} \cdot \vec{p}
		\;\mp\; \hat{n} \cdot \vec{p} \;\gamma_o \,\right)
		\,g_{_{2}} 			\label{eq-G-rep}
\end{eqnarray}
	where the scalar functions $f, g$ are even under the exchange
	$\hat{n} \rightarrow - \hat{n}$.
	This representation ensures that the thermal fermion propagator
	is spacetime translation invariant.  This arises because
	according to eq.(\ref{eq-psi-Psi-exp}), the thermal fermion Green
	function may be written in terms of the canonical field
	$\Psi$ and the new spacelike $\Psi^{s}$
\begin{eqnarray}
	< \psi (x) \bar{\psi} (y) > &=& < \Psi (\x)
			\overline{\Psi} (\y) > \nonumber \\
	& &	- \frac{\Tprime}{4\;\;} \;\int dt' \,\eps(y_o - t') \;
		< \Psi (\x) \overline{\Psi} (\vec{y} -
			\hat{n} (y_o - t'), t') \; \gnplusdef > \nonumber \\
	& &	+ \frac{\Tprime}{4\;\;} \;\int dt' \,\eps(t-t') \;
		< \gnplusdef \;\Psi ( \vec{x} - \hat{n} (t-t'), t')
			\;\overline{\Psi} (\y) >  \nonumber \\
	& &	+ <\Psi (\x) \overline{\Psi^{s}} (\y) >
		+ <\Psi^{s} (\x) \overline{\Psi} (\y) >
		+ <\Psi^{s} (\x) \overline{\Psi^{s}} (\y) > \nonumber \\
\end{eqnarray}
	Here, for simplicity, we have identified the two lightcone
	averages and written them as a single average over the orientation,
	$\hat{n}$.  The first three terms in the expansion involve
	only the canonical Dirac field $\Psi$, and, as pointed out in the
	earlier work $\cite{Chang-xc}$, they {\em together
	correctly reproduce the spinor as well as the pole
	structure } of the thermal Green function.  The chiral
	flip part of the canonical massive Green function $< \Psi
	\overline{\Psi} >$ is cancelled by the other two terms, thus
	agreeing with the apparent chiral symmetry of the thermal
	Green function $ < \psi \bar{\psi} >$.

	The cross-terms involving $\Psi$ with
	$\Psi^{s}$ would, however, give rise to Fourier integrals of the type
\[
	{\rm e}^{i \vec{p} \cdot (\vec{x} - \vec{y})}
	{\rm e}^{- i p_o t \;\pm\;  i\, \hat{n} \cdot \vec{p} \,y_o}
\]
	which are manifestly not invariant under time translations.
	The Dirac matrix representation in eq.(\ref{eq-F-rep},
	\ref{eq-G-rep}) ensures the absence of these terms.
	The last term in the expansion reproduces correctly the
	spacelike cuts of the thermal Green function, when
\begin{eqnarray}
	F_{_{+}} &=&   \frac{\eta}{2} \;\left\{ \;\;\;
			\frac{\gnminusdef}{2} + i
		\frac{ (\vec{\gamma}\cdot \vec{p} \;-\; \hat{n} \cdot
		\vec{p} \,\gamma_o) }{\sqrt{ p^2 - (\hat{n} \cdot \vec{p})^2}
		     }   \right\} 		\label{eq-F-plus} \\
	F_{_{-}} &=&   \frac{\eta}{2} \;\left\{
			- \frac{\gnplusdef}{2} - i
		\frac{ (\vec{\gamma}\cdot \vec{p} \;+\; \hat{n} \cdot
		\vec{p} \,\gamma_o) }{\sqrt{ p^2 - (\hat{n} \cdot \vec{p})^2}
		     }    \right\}		\label{eq-F-minus} \\
	G_{_{+}} &=&   \frac{\eta'}{2} \;\left\{ \;\;\;
			\frac{\gnminusdef}{2} + i
		\frac{ (\vec{\gamma}\cdot \vec{p} \;-\; \hat{n} \cdot
		\vec{p} \,\gamma_o) }{\sqrt{ p^2 - (\hat{n} \cdot \vec{p})^2}
		     }   \right\} 		\label{eq-G-plus} \\
	G_{_{-}} &=&   \frac{\eta'}{2} \;\left\{
			- \frac{\gnplusdef}{2} - i
		\frac{ (\vec{\gamma}\cdot \vec{p} \;+\; \hat{n} \cdot
		\vec{p} \,\gamma_o) }{\sqrt{ p^2 - (\hat{n} \cdot \vec{p})^2}
		     }    \right\}		\label{eq-G-minus}
\end{eqnarray}
	Here $\eta , \eta'$ are overall phase factors that remain
	arbitrary.
	The requirement that the
	$< \Psi^{s} \overline{\Psi^{s}} >$ reproduce
	the cut structure exhibited in eq.(\ref{eq-spacelike-cut})
	only demands an internal $90^{o}$ phase difference between
	$f_{_{1}}$ and $f_{_{2}}$, and likewise between $g_{_{1}}$
	and $g_{_{2}}$.
	The choice for the overall phase factors in $F_{_{\pm}}$
	and $G_{_{\pm}}$ will become apparent only when we turn to
	the relation between $\psi$ and the canonical Dirac field $\Psi$
	at time $t=0$ and study their connection with the thermal
	vacuum at high $T$.

\section{Relation to Thermal Vacuum}

	Eq.(\ref{eq-psi-Psi-exp}) provides the expansion of
	$\psi (x)$ in terms of the canonical field, $\Psi (x)$.
	In this section, we shall use it to relate the $\psi$
	field at $t=0$ to the canonical operators.  Let
\begin{equation}
	\psi(\vec{x}, 0) =  	\frac{1}{\sqrt{V}} \sum_{p} \;
				{\rm e}^{i \vec{p} \cdot \vec{x}} \;
			\left\{  \left( \begin{array}{c}
					\chi_{_{L}} \apL \\
					\chi_{_{R}} \apR
					\end{array} \right)
				+ \left( \begin{array}{c}
					\;\;\chi_{_{R}} \bdagR \\
					- \chi_{_{L}} \bdagL
					\end{array} \right)
			 \right\}
\end{equation}
	where $\chi_{_{L,R}}$ is an eigenfunction of the helicity operator
	$ \vec{\sigma} \cdot \hat{p}$ with eigenvalue $\pm 1$ respectively.
	A straightforward evaluation of the gamma matrix algebra
	implied in eq.(\ref{eq-psi-Psi-exp}) shows that
\begin{eqnarray}
	\aps	    &=& \Aps \;+\; \eta' \;s\; \frac{\Tprime}{2 p}  \Bdag
			+ O(\Tprimesq /p^2) 	\label{eq-aps-Aps} \\
	b^{}_{p,s}  &=& B^{}_{p,s} \;-\; \eta^{*} \;s\; \frac{\Tprime}{2 p}
			A^{\dagger}_{-p,s}
			+ O(\Tprimesq /p^2)	\label{eq-bps-Bps}
\end{eqnarray}
	At this stage, the rule that $\aps$ and $b^{}_{p,s}$ should
	anticommute, places the requirement
\begin{equation}
	\eta'   \;=\;   \eta^{*}
\end{equation}
	leaving us with still an overall phase factor $\eta$ to
	be determined.

	In the absence of $\Psi^{s}$, the relation between $\aps,
	b^{}_{p,s}$ and $\Aps, B^{}_{p,s}$ would have been simply
	$\aps = \Aps$ and $b^{}_{p,s} = B^{}_{p,s}$.
	The additional terms of $\Bdag$ and $A^{\dagger}_{-p,s}$ arose
	entirely from the spacelike $\Psi^{s}$ field in
	eq.(\ref{eq-psi-Psi-exp}).

	{\em  Thus the spacelike cut in the fermion propagator signals the
	presence of Bogoliubov pairing in the thermal vacuum.}

	Can we determine the precise nature of this Bogoliubov pairing
	in the new thermal vacuum?

	The answer is yes. For a study of the order parameter
	$\psibarpsi_{_{\beta}}$ will yield information on the properties
	of the vacuum.
	Since we now have the relation between $\psi$ and $\Psi$ at time
	$t=0$, we may proceed to obtain the canonical expansion for
	$\bar{\psi} \psi$.
	Let us introduce the order operator$\cite{Chang-chiralg}$
\begin{equation}
	\frac{1}{2} \int d^3 x  \bar{\psi} (\vec{x}, 0) \psi (\vec{x}, 0)
		=  - \sum_{p} Y_{_{1p}}
\end{equation}
	where
\begin{equation}
	Y_{_{1p}} \equiv - \sum_{s} \frac{s}{2}  \left( a^{\dagger}_{p,s}
		b^{\dagger}_{-p,s}
		+ a^{}_{p,s} b^{}_{-p,s} \right)	\label{eq-order-op}
\end{equation}
	The order parameter $< \bar{\psi} \psi>_{_{\beta}}$ is the
	expectation value of the order operator
	with respect to the new thermal vacuum $| vac >$.

	Recall that the original Fock space vacuum satisfies the property
\begin{equation}
	\aps  \;| 0 >   \;=\;  b^{}_{p,s} \;|0> \;=\; 0
\end{equation}
	while the new thermal vacuum is the one annihilated by $\Aps,
	\Bps$
\begin{equation}
	\Aps  \;| vac >   \;=\;  \Bps \;|vac> \;=\; 0
\end{equation}
	Upon substituting the relations eq.(\ref{eq-aps-Aps},
	\ref{eq-bps-Bps}) into the order operator eq.(\ref{eq-order-op}),
	we find the canonical expansion
\begin{equation}
	\int d^3 x \bar{\psi} (\vec{x}, 0) \psi (\vec{x}, 0)
	= \sum_{p,s} \left\{ s \left( \Adag \Bdag
		+ \Bps \Aps \right) - (\eta + \eta^{*}) \frac{\Tprime}{2p}
		\left( \Adag \Aps - \Bps \Bdag \right)
		\right\}
\end{equation}
	and we have the order parameter at $t=0$
\begin{eqnarray}
	\int d^3 x < \bar{\psi} \psi >_{_{\beta}}  &=&
		\sum_{p,s} \; < vac |
		\left\{ s \left( \Adag \Bdag
		+ \Bps \Aps \right) - (\eta + \eta^{*}) \frac{\Tprime}{2p}
		\left( \Adag \Aps - \Bps \Bdag \right)
		\right\} |vac>  	\nonumber \\
	&=&	(\eta + \eta^{*}) \;\sum_{p} \frac{\Tprime}{p}
\end{eqnarray}
	Since $< \bar{\psi} \psi >_{_{\beta}}$ vanishes, we finally arrive
	at the result that\footnote{
\samepage \sl There is, of course, a conjugate solution with
	$\eta = -\eta' = -i$.  To the extent that they yield the same
	spacelike cut contribution, it is physically equivalent to
	the solution we pick.
}
\begin{equation}
	\eta  = - \eta'=   i
\end{equation}
	which in turn implies that to order $O(\Tprime / p)$, we have
\begin{eqnarray}
	\aps	&=& \Aps \;-\; i \;s\; \frac{\Tprime}{2 p}
			\Bdag 			\label{eq-aps-Aps-1} \\
	b^{}_{p,s}	&=& B^{}_{p,s} \;+\; i \;s\; \frac{\Tprime}{2 p}
			A^{\dagger}_{-p,s} 	\label{eq-bps-Bps-1}
\end{eqnarray}
	This relation spells out the precise nature of the Bogoliubov
	tranformation that takes us from the original Fock space
	vacuum to the new thermal vacuum.  The new vacuum is the
	generalized Nambu-Jona-Lasinio vacuum$\cite{NJL}$
\begin{equation}
	| vac >  =  \prod_{p,s} \left( \costhetap \;-\; i \;s
			\;\sinthetap \;\adagbdag \right) \; | 0 >
\end{equation}
	where
\begin{equation}
	\tan{ 2 \thetap} = \frac{\Tprime}{p}
\end{equation}
	The interesting feature of this generalized NJL vacuum is the
	presence of the phase factor $i$ in the quark-antiquark pair.

	As has been pointed out elsewhere$\cite{Chang-xc,Chang-Banff}$
	this phase factor is responsible for the vanishing of $\psibarpsi$,
	without however breaking up the quark-antiquark pairs.  As a result
	the generalized NJL vacuum is not chiral-invariant. By this, we mean
	that under the old (zero temperature) chirality
\begin{eqnarray}
	\aps  &\rightarrow& {\rm e}^{i s \,\alpha} \aps  \\
	\bps  &\rightarrow& {\rm e}^{i s \,\alpha} \bps
\end{eqnarray}
	the thermal vacuum goes over into a a new unitarily inequivalent
	vacuum.

	And yet there is an apparent chiral symmetry of the BPFTW action
	in eq.(\ref{eq-BP-action}).  The resolution of this paradox
	comes when we recognize that the new chiral symmetry at
	high $T$ is associated with a different Noether
	charge$\cite{Weldon-BP}$ than the original (zero temperature)
	chirality.  This new
	$\Qbeta$ has the expansion$\cite{Chang-xc}$
\begin{equation}
	\Qbeta = - \frac{1}{2}  \sum_{p,s} s \left( \Adag \Aps
			+ \Bdag \Bps \right)
\end{equation}
	which clearly annihilates the thermal vacuum $|vac>$ and explains
	the apparent chiral symmetry of the high temperature effective
	action.
	But this high temperature chirality is not identical to
	the old chirality, $\Q$ given by
\begin{equation}
	\Q  =  - \frac{1}{2} \sum_{p,s} s \left( \adag \aps
			+ \bdag \bps \right)
\end{equation}

\section{Conclusion}
	In this paper, we have performed the quantization of the
	BPFTW action in coordinate space.  By reformulating it
	in terms of auxiliary fields, the action may be made
	local.  The auxiliary fields, $\chi_{_{\pm}}$, satisfy
	local equations of motion that admit homogeneous solutions,
	which turn out to be directly related to the spacelike
	cuts in the thermal propagator.

	The homogeneous solutions impact on the relation
	between $\psi$ and the canonical $\Psi$ field at $t=0$.
	These relations spell out the connection between
	the vacuum of the massless free field $\aps, \bps$
	and the thermal vacuum of the effective action.
	Our results show how the presence of the spacelike cuts
	in the fermion propagator signals a new phase
	in the underlying thermal vacuum.

	This thermal vacuum continues to exhibit
	the rich and complex structure of chiral symmetry violations as the
	zero temperature case.  That $\psibarpsi$ vanishes is no
	proof of chiral restoration at high $T$.
	In a separate communication$\cite{Chang-chiralg}$, we introduce
	and discuss the $SU(2N_f)_{p} \otimes SU(2N_f)_{p}$ chirality
	algebra of order parameters that probe the state the chiral
	symmetry breaking.

	Many questions remain to be answered.  What is the underlying
	physics origin of the auxiliary field?  How do the quark-antiquark
	pairing at high temperatures generate the intriguing $90^{o}$
	phase in the BCS-like ground state?  There is clearly
	an interplay between the new chirality $\Qbeta$ at high $T$
	and the old $\Q$ chirality.  How does it impact on the
	interactions of the pion at high $T$?$\cite{Chang-QCD-pion}$



\begin{thebibliography}{99}
\bibitem{notation}
	We use the metric such that $a \cdot b \equiv
	\vec{a} \cdot \vec{b} - a_o b_o$.  Here
	$\gamma_{\mu} = (\vec{\gamma}, \gamma_o )$, with
	$\vec{\gamma}$ hermitian, while $\gamma_o$ is
	antihermitian.
\end{thebibliography}
\end{document}